\documentclass[a4paper,usenatbib]{mnras}
\usepackage{newtxtext,newtxmath}
\usepackage[T1]{fontenc}
\usepackage{ae,aecompl}

\usepackage{graphicx}	% Including figure files
\usepackage{amsmath}	% Advanced maths commands
\usepackage{fontawesome}

\newcommand{\nblink}[1]{\href{https://github.com/DifferentiableUniverseInitiative/DHOD/blob/master/nb/#1.ipynb}{\faFileCodeO}}
\newcommand{\github}{\href{https://github.com/DifferentiableUniverseInitiative/DHOD}{\faGithub}}

\newcommand{\dhod}{{\sc DiffHOD}}
\title[DiffHOD]{Differentiable Stochastic Halo Occupation Distribution}

\author[B. Horowitz et al.]{
Benjamin Horowitz,$^{1,2}$\thanks{E-mail: bhorowitz@princeton.edu}
ChangHoon Hahn,$^{1}$
Francois Lanusse,$^{3}$
Chirag Modi,$^{4,5}$ \newauthor Simone Ferraro$^{2}$
\\
$^{1}$Department of Astrophysical Sciences, Princeton University, Princeton, NJ 08544, USA\\
$^{2}$Lawrence Berkeley National Lab, 1 Cyclotron Road, Berkeley, CA 94720, USA\\
$^{3}$AIM, CEA, CNRS, Universit\'e Paris-Saclay, Universit\'e Paris Diderot, Sorbonne Paris Cit\'e, F-91191 Gif-sur-Yvette, France\\
$^{4}$Center for Computational Astrophysics, Flatiron Institute, 162 Fifth Ave., New York, NY 10010, USA\\
$^{5}$Center for Computational Mathematics, Flatiron Institute, 162 Fifth Ave., New York, NY 10010, USA
}

\date{Accepted XXX. Received YYY; in original form ZZZ}

\pubyear{2021}

\begin{document}
\label{firstpage}
\pagerange{\pageref{firstpage}--\pageref{lastpage}}
\maketitle

\begin{abstract}
In this work, we demonstrate how differentiable stochastic sampling techniques developed in the context of deep Reinforcement Learning can be used to perform efficient parameter inference over stochastic, simulation-based, forward models. As a particular example, we focus on the problem of estimating parameters of Halo Occupancy Distribution (HOD) models which are used to connect galaxies with their dark matter halos. Using a combination of continuous relaxation and gradient re-parameterisation techniques, we can obtain well-defined gradients with respect to HOD parameters through discrete galaxy catalogs realisations. Having access to these gradients allows us to leverage efficient sampling schemes, such as Hamiltonian Monte-Carlo, and greatly speed up parameter inference.
%Using the Gumbel-Softmax approach, we can map the discrete HOD models to a continuous distribution with well defined derivatives, allowing the use of first order optimization methods, such as Hamiltonian Monte Carlo, for analysis of data.
We demonstrate our technique on a mock galaxy catalog generated from the Bolshoi simulation using the 
\cite{2007zheng} %Zheng et al. 2007 
HOD model and find %,finding 
near identical posteriors as standard Markov Chain Monte Carlo techniques with an increase of $\sim 8$x in convergence efficiency. 
Our differentiable HOD model %This model 
also has broad applications in full forward model approaches to cosmic structure and cosmological analysis. \github
\end{abstract}


%%%%%%%%%%%%%%%%%%%%%%%%%%%%%%%%%%%%%%%%%%%%%%%%%%

%%%%%%%%%%%%%%%%% BODY OF PAPER %%%%%%%%%%%%%%%%%%

\section{Introduction}

%plot comparing DiffHOD to HOD at PS level

%MCMC run with HOD parameters

Over the past twenty years, there has been significant observational and theoretical progress in connecting galaxies to their cosmic environments \citep{2000MNRAS.318.1144P,2003MNRAS.340..771V,kravtsov2004,2006ApJ...652...71W,2011MNRAS.414.1405N}. Understanding this connection is critical for understanding galaxy formation/evolution \citep{2009MNRAS.399.1773C,2011ApJ...736...59Z} as well as using galaxies as bias tracers of the underlying mass density for cosmological analyses \citep{2000MNRAS.311..793B,2018PhR...733....1D}. Studying this connection is a key component of many upcoming galaxy surveys including the Prime Focus Spectrograph \citep{takada2014, 2016SPIE.9908E..1MT}, the Dark Energy Spectroscopic Instrument \citep{2016arXiv161100036D, abareshi2022}, and the Nancy Grace Roman Space Telescope~\citep{2015arXiv150303757S, wang2022_roman}.

A key theoretical tool for these studies has been the Halo Occupation 
Distribution \citep[HOD;][]{lemson1999, benson2000, seljak2000, scoccimarro2001, 2002ApJ...575..587B, wechsler2018}, 
a framework that specifies how collapsed dark matter halos~\citep{press1974, bond1991, cooray2002} are populated with galaxies.
 %an abundance matching approach to map from dark matter halos to their constituent galaxies. 
This is in contrast to ``environmental" biasing schemes, such as Eulerian 
or Lagrangian biasing schemes \citep{1998MNRAS.293..209M, 2018PhR...733....1D}, 
common in 
%the analysis of galaxy survey data for cosmological analysis 
cosmological analyses of galaxy survey data \citep{2016MNRAS.457.1770C,2020JCAP...05..042I,2021PhRvL.126n1301T, beutler2017}.
Unlike environmental biasing schemes that %matches only models the summary statistics 
only model summary statistics
like power spectrum of the galaxy field,
a well-formulated HOD model provides direct physical insight into galaxy formation physics 
through its parameters, which are %as the parameters of an HOD 
related to critical mass scales in galaxy-halo relation. 
For example, this allows direct measurement of HOD parameters by comparing observed galaxy populations with dynamical mass measurements, such as x-ray clusters \citep{2009ApJ...707..554Z,2016MNRAS.463.1929M}. 
%\CH{fine --- but the physical meaningfulness of HOD is a bit overstated here}

In standard HOD implementations~\citep[\emph{e.g}][]{2007zheng},  %, such as that in the standard halotools implementation \citep{2016MNRAS.460.2552H}
the HOD model specifies the probability distribution of the number
of galaxies, $N$, hosted by a dark matter halo given its properties, such as halo 
mass: $P(N|M_{\rm halo})$. A semi-analytical halo-model approach can include HOD parameters to predict two and higher point function \citep{cooray2002}, but those predictions are often not accurate 
enough for analysing modern datasets. Alternatively, a more precise Monte Carlo approach is often used to stochastically assign galaxies to halos in a large simulation box following the HOD prescription, and then the galaxy power spectrum or other quantities of interest are directly measured from the simulation. 
% Formally this assignment is not differentiable as it is over a discrete categorical variable. 
% This greatly restricts the possible approaches for parameter fitting and error estimation. 

In practice, Markov-Chain Monte-Carlo methods have primarily been 
used to fit HOD parameters from mock or actual data~\citep[\emph{e.g.}][]{white2011, rodriguez-torres2016, sinha2018}. However these methods scale poorly with the number of parameters that need to be fit.
As novel decorated HOD models increasingly add more assembly bias parameters to accurately capture small scale observations,
this exercise can become challenging, especially if there are unforeseen degeneracies in the parameter space.
These challenges can be overcome more easily with parameter inference methods that rely on the gradient information 
\emph{i.e.} where we can estimate the response of the observations with respect to the underlying parameters of the model,
such as Hamiltonian Monte Carlo \citep{1987PhLB..195..216D}, Variational Inference \citep{peterson1987mean,2003PhDT.......250B,2016arXiv160100670B} or combinations thereof
\citep{gabrie22, Modi22}.
However these methods are not applicable in current HOD models as the implementation of their stochastic galaxy assignment schemes make them classically non-differentiable. 

A differentiable HOD framework will enhance dynamical forward 
modelling frameworks that seek to reconstruct latent cosmological 
fields~\citep[\emph{e.g.}][]{2017JCAP...12..009S}, which are
constrained to use gradient based methods for optimization due to the high dimensionality of the inference problem.
These frameworks generally rely on perturbative bias models that are accurate only on large scales \citep{Schmidt18, Modi19} 
or heuristic neural network models with a large number of latent parameters \citep{2018JCAP...10..028M}.
A differentiable HOD approach will allow one to push to smaller scales with a well-understood, physically developed model that has only a handful of parameters.
%\CH{we need a sentence or two on why we care about these gradient-based methods. perhaps provide some bigger picture context of DUI.} 

An alternative to HOD models that maintains the requisite differentiability is to use differentiable emulators \citep{2015ApJ...810...35K,2020MNRAS.492.2872W} or fitting functions of the observables like the one proposed in \citet{2021arXiv210505859H} for galaxy assembly bias. 
However these are efficient only for the particular summary statistics and cosmological parameters on which they are trained.
Hence, they require a new training set once these are varied. 
Depending on the parameter space of interest this could be of prohibitive computationally cost. 
In addition, separate emulators must be trained for each summary statistic of interest as such methods do not match the galaxy observations at the field level.

%\FL{Other emulators/fitting functions used for HMC that we can cite? We also want to say why they are annoying, like require a "training set" of simulations, not flexible, etc..}

Motivated by this, here we adopt a different approach and aim to make the HOD sampling itself differentiable.
Our aim is to be able to compute gradients of any observable with respect to HOD parameters through a particular realisation of a galaxy catalog. 
Common wisdom states that differentiating through stochastically sampled discrete random variables, such as the number of satellites in a given halo, is not possible. 
However, modern Reinforcement Learning have spurred the development of techniques to deal with these types of categorical variables in the context of deep neural network training via back propagation. 
In particular, we use the Gumbel-Softmax or CONCRETE method \citep{2016arXiv161101144J,2016arXiv161100712M} 
that utilises continuous distributions to approximate the sampling process of discrete stochastic variables, such as galaxies, in a differentiable fashion. %\FL{this last sentence needs a little bit more work.}
It relies on two insights: 1) a re-parameterization for a discrete (or categorical) 
distribution in terms of the Gumbel distribution~\citep[referred to as the ``Gumbel trick'';][]{2014arXiv1411.0030M} 
and 2) making the corresponding function continuous by using a continuous approximation that depends on a \textit{temperature} parameter, which in the zero-temperature case degenerates to the discontinuous, original expression.

In this paper, we will implement the Gumbel-Softmax method in the context of HOD models and apply it to mock datasets. In Sec. \ref{sec:methods} we will describe our HOD model and the methods used to allow differentiability of its categorical outputs. In Sec. \ref{sec:exp}, we apply this technique to a Monte Carlo analysis of a mock galaxy catalog constructed from the Planck-Bolshoi simulation. In Sec. \ref{sec:conclusion} we compare the differentiable HOD model to that from a standard approach and discuss its applications.

\section{Method}
\label{sec:methods}

In this section, we provide some background on the various components of our HOD model, and detail our strategy to make this model differentiable. We implement our model using Tensorflow Probability \citep{2016arXiv161009787T,morgan2018probabilistic}, particularly the Tensorflow Distribution package \citep{2017arXiv171110604D}. 

\subsection{HOD Model} \label{sec:hod}

To describe the population of galaxies in our halos we use the standard 
\cite{2007zheng} HOD model.
In the \cite{2007zheng} model, the probability of a given halo hosting 
$N$ galaxies is dictated solely by its mass --- $P(N|M)$.
The model separately populates central and satellite galaxies, motivated by
theoretical studies~\citep{kravtsov2004, 2005zheng}, and has five free parameters
with some physical significance that
can be related back to well studied mass-luminosity relationships.

\subsubsection{Central occupation}

For central galaxies, the mean occupation function is step-like with a soft cutoff to account for natural scatter between galaxy luminosity the halo host mass. There are two free parameters controlling this function, the characteristic minimum mass of halos hosting central galaxies above some luminosity threshold, $M_\textrm{min}$, and the width of the cutoff profile, $\sigma_{\log M}$:
\begin{equation}
    \langle N_\textrm{cen}(M) \rangle = \frac{1}{2} \left[ 1 + \textrm{erf}\left( \frac{\log M - \log M_\textrm{min}}{\sigma_{\log M}}\right)\right],
    \label{eq:ncen}
\end{equation}
where erf is the standard error function and $M$ is the halo mass. 
Given the mean occupation for a halo of a given mass, central galaxies
are assigned to halos by sampling a Bernoulli distribution:
\begin{equation}
    N_\textrm{cen} \sim \textrm{Bernoulli}\big( p = \langle N_\textrm{cen}(M) \rangle \big).
    \label{eq:ncensample}
\end{equation}
%\CH{explain how $N_{\rm cen}$ is sampled for each halo} 

\subsubsection{Satellite occupation}

Simulations suggest that satellites follow an approximately power-law distribution with a slope close to unity at the high mass end. At lower masses, the shape of the distribution changes and the overall distribution can be parameterized as
%\langle N_\textrm{sat}(M) \rangle = \frac{1}{2} \left[ 1 + \textrm{erf}\left( \frac{\log M - \log M_\textrm{min}}{\sigma_{\log M}}\right)\right]\left( \frac{M-M_0}{M_1'}, \right)^\alpha,
\begin{equation}
    \langle N_\textrm{sat}(M) \rangle = \langle N_{\rm cen}(M) \rangle \left( \frac{M-M_0}{M_1'} \right)^\alpha,
    \label{eq:nsat}
\end{equation}
where $\alpha$ is the power law slope at high masses, $M_0$ is the characteristic mass of the change-over and $M_1'$ is the characteristic amplitude.  This mean number of satellites for a given mass is then used to define the intensity $\lambda$ of a Poisson distribution, from which a particular number of satellites are drawn for each halo.
\begin{equation}
    N_\textrm{sat} \sim \textrm{Poisson}\big(\lambda = \langle N_\textrm{sat}(M) \rangle \big)
    \label{eq:nsatsample}
\end{equation}
%hodpy \cite{2017MNRAS.470.4646S}, 

%\CM{Do we worry about the fact that some models do not have satellites in the halos that do not have any centrals?}

\subsubsection{Spatial satellite distribution}

In the \cite{2007zheng} HOD, central galaxies are located at the center 
of its host halo and satellite galaxies are distributed according to a 
\cite{nfw1997} profile (hereafter NFW).
To sample the satellite galaxy positions, we utilize the \cite{2018RNAAS...2...55R} 
implementation, which constructs an efficient mapping from a random sample 
and the full NFW profile via the calculation of the quantile function, 
\emph{i.e.} the inverse Cumulative Distribution Function (CDF). 
This can be written analytically as

\begin{equation}
    q(p;c,M_{\rm vir}) = -\frac{1}{c}\left[ 1 + \frac{1}{W_0(-e^{-pM_{\rm vir}-1})} \right],
    \label{eq:NFW}
\end{equation}
where $W_0$ is the Lambert-W function, $M_{\rm vir}$ is the virial mass and $c$ is the concentration parameter. Using this inverse CDF, we can now randomly draw radial distances for satellites by sampling  $p$ from $U[0,1]$ and mapping it to radii as $r/r_{\rm vir} = q(p,c)$. The angular position of the satellite is sampled uniformly in this isotropic model.

\subsection{Differentiable stochastic sampling}

In this section, we review the key ideas behind differentiable stochastic sampling, which will form the building blocks of DiffHOD.

\subsubsection{Stochastic backpropagation by reparametrisation}

One of the most common approaches for backpropagation through stochastic sampling is the so-called reparametrisation trick, extensively used for instance in Variational Auto-Encoders \citep{Kingma2013,Rezende2014}. 

The key idea of this approach is to rewrite samples $z$ from a given paramteric distribution $\mathbb{P}_{\theta}$ as a deterministic and differentiable transformation $f$ applied to a fixed distribution $\mathbb{P}_{\epsilon}$:
\begin{equation}
    z = f(\theta, \epsilon) \mbox{ with } \epsilon \sim \mathbb{P}_{\epsilon} 
\end{equation}
This reparametrisation of the samples allows to side-step having to take derivatives of the stochastic variable $\epsilon$ when computing derivatives of some downstream function $h$ with respect to the distribution parameters $\theta$:
\begin{equation}
    \frac{\partial}{\partial \theta} \mathbb{E}_{z \sim p_\theta}\left[ h(z) \right] = 
    \mathbb{E}_{\epsilon \sim p_\epsilon} \left[ \frac{\partial}{\partial \theta} h(f(\theta, \epsilon)) \right] %= \mathbb{E}_{\epsilon \sim p_\epsilon} \left[ \frac{\partial h}{\partial f}  \frac{\partial f}{\partial \theta} \right]
\end{equation}
In the right-hand side of this expression, the derivative now only involves taking gradients of a deterministic function of $\theta$. 

To provide a simple concrete example of such reparameterisation, let us consider a Gaussian distribution of mean $\mu$ and standard deviation $\sigma$. One can express a sample $z \sim \mathcal{N}(\mu, \sigma^2)$ as $z = \mu + \sigma \epsilon$ with $\epsilon \sim \mathcal{N}(0, I)$, making it trivial to take derivatives of the samples with respect to the parameters of the distribution ($\mu$ and $\sigma$).

\subsubsection{Gumbel-Softmax trick for categorical variables}

The reparameterisation trick as presented above requires the samples to be expressible as a deterministic and differentiable function of a random variable. While this can often be achieved for continuous distributions, it is typically not directly possible for discrete categorical variables. 
To overcome this limitation, the Gumbel-Softmax trick \citep{2016arXiv161101144J,2016arXiv161100712M} introduces a relaxation of a categorical distribution to a continuous distribution, which can then be handled with the reparameterisation trick.

%We review here the Gumbel-Softmax trick as developed in the literature \citep{2016arXiv161101144J,2016arXiv161100712M} to map from categorical variables to a continuous differentiable distribution. 
Let $z$ be a categorical variable with class probabilities $\pi_1,\pi_2,...\pi_j$ that we wish to sample.
We assume that categorical samples are encoded as $N$-dimensional one-hot vectors, 
\emph{i.e.} they are 1$\times N$ vectors with all elements 0 except the
the element corresponding  to the sampled class which is 1. 
The simplest way to sample z is by 
\begin{equation}
    z = \textrm{onehot}(\textrm{max}{\, i | \pi_1 + ... + \pi_{i-1}\leq U}), \quad U \sim \textrm{Uniform}(0, 1)
    \label{eq:naivesampling}
\end{equation} 
A first step towards making these samples differentiable is to use the Gumbel-Max trick \citep{gumbel1954statistical,2014arXiv1411.0030M}, which reparametrizes categorical sampling as 
\begin{equation}
    z = \textrm{onehot}\left(\textrm{argmax}_{i}[g_i + \log{(\pi_i)}]\right),
    \label{eq:reparametrize}
\end{equation}
where $g_i$ are i.i.d. random variables drawn from the Gumbel distribution between 0 and 1, $\textrm{Gumbel}(0, 1)$\footnote{The standard Gumbel distribution is defined by cumulative distribution function $\textrm{CDF}(x) = \exp(-\exp(x))$ and probability density function, $\textrm{PDF}(x) = \exp(-(x+\exp(-x))$.}. 
This reparameterization trick refactors the sampling of $z$ into a deterministic function of the parameters ($\pi$) and some independent noise with a fixed distribution.
%As a result, now gradients are propagated w.r.t. parameters of a deterministic function instead of computing the gradient w.r.t. parameters of a distribution in Eq. \ref{eq:naivesampling}, which is a harder problem.

However the reparametrized function is still non-differentiable due to the $\textrm{argmax}$ function.
A continuous, differentiable approximation to this is given by a softmax function, 
\begin{equation}
    \mathrm{softmax}(\mathbf{z}, \tau )_i = \frac{\mathrm{e}^{z_i/\tau}}{\sum_{j=1}^k \mathrm{e}^{z_j/\tau}}, \quad \mathbf{z} = (z_1, z_2,...,z_k)
 \end{equation}
where $\tau$ is a free parameter sometimes referred to as the ``temperature".

Using this approximation relaxes the discreteness of the Gumbel-Max trick and generates a $k$-dimensional vector $\mathbf{z}$ 
\begin{equation}
    \hat{z}_i = \frac{\exp((\log(\pi_i)+g_i)/\tau)}{\sum_j \exp((\log(\pi_j)+g_j)/\tau)},
    \label{eq:softmax}
\end{equation}
We recover the true discrete function in the limit of $\tau \rightarrow 0$.
As this function is analytical in class probabilities $\pi$ for $\tau > 0$,
we can estimate the gradients of the observed samples $\mathbf{z}$ with respect to the parameters parameterizing $\pi$.

In the remaining of this work, we will make use of the special case when the number of classes is 2, i.e. when the categorical distribution reduces to a Bernoulli distribution. In this special binary case Eq. \ref{eq:softmax} can be simplified. \cite{2016arXiv161100712M} refer to the resulting distribution as BinConcrete; we will refer to it in this work as a Relaxed Bernoulli distribution. Using the fact that the difference of two Gumbel variables follows a Logistic distribution\footnote{The standard Logistic distribution follows the following probability density function: $p(x) = \frac{\exp(-x)}{(1 -\exp(-x))^2}$}, \citep{2016arXiv161100712M} shows the Relaxed Bernoulli can be reparameterized as:
\begin{equation}
    z = \frac{1}{1 + \exp( - (\log \pi + \epsilon)/\tau) } \mbox{ with } \epsilon \sim \mathrm{Logistic}(0,1) \;.
\end{equation}
where in this expression $\pi$ is the odds-ratio $\pi=p/(1 -p)$ if $p$ is the probability of corresponding Bernoulli distribution.

\subsection{\dhod~Implementation} \label{sec:dhod}

We now have all the elements needed to build a differentiable HOD (hereafter \dhod) model. 
We describe in this section our strategy for sampling central and satellites occupation, and satellites positions.

\subsubsection{Differentiable central occupation sampling}

As described in Sec.~\ref{sec:hod}, the central occupation is defined by a Bernoulli distribution, with a parameter $p=\left\langle N_{\rm cen}(M | M_\mathrm{min}, \sigma_{\log M})\right\rangle$ defined as a deterministic function. 

Here, we can directly apply the Gumbel-Softmax trick introduced above in the specical case of a binary variable. We therefore sample the central occupation of each halo using the Relaxed Bernoulli distribution:
\begin{equation}\label{eq:diff_cen}
 N_{\rm cen} = \frac{1}{1 + \exp( - (\log(\frac{p}{1 - p}) + \epsilon)/\tau) } \mbox{ with } \epsilon \sim \mathrm{Logistic}(0,1) \;.   
\end{equation}
at given temperature $\tau$.
Fig.~\ref{fig:occupancy} illustrates the high agreement between the halo occupancy obtained by sampling centrals with this Relaxed Distribution compared to the analytical expectation at our fiducial choice of $\tau=0.1$. 

\subsubsection{Differentiable satellite occupation sampling}

For satellites, we aim to define a differentiable approach to sampling from a Poisson distribution with intensity $\lambda=\left\langle N_{\rm sat}(M | M_0, M_1^\prime, \alpha)\right\rangle$, also a deterministic and differentiable function. To build on the Gumbel-Softmax trick, we propose to replace conventional Poisson sampling of the total number of satellites by sampling each satellite individually from a Bernoulli distribution. 

Let us consider a halo with a Poisson rate $\lambda$ for its satellite occupancy. We assume the halo can have a maximum of $N$ satellites, then for each potential satellite we sample from a Bernoulli distribution with probability $p=\lambda/N$ whether this satellite will be included in the halo. The resulting statistics of the number of satellites with this procedure will be Binomial (as N draws from i.i.d. Bernoulli). 

More formally, we propose to approximate the Poisson distribution with intensity $\lambda$ of a standard HOD by a Binomial distribution with $N$ trials and probability $p=\lambda/N$:
\begin{equation}
    N_{\rm sat} \sim \mathrm{Binomial}\left(N, p=\frac{ \left\langle N_{\rm sat}\right\rangle }{N} \right)
\end{equation}
By construction, this Binomial distribution will yield the same mean number of satellites as the Poisson distribution, however the variance of both distributions is different:
\begin{align}
    \mathrm{Var}(N_{\rm sat}^{\rm Pois.}) &= \left\langle N_{\rm sat}\right\rangle \\  \mathrm{Var}(N_{\rm sat}^{\rm Bin.}) &= \left\langle N_{\rm sat}\right\rangle * \left(1 - \frac{\left\langle N_{sat}\right\rangle}{N}\right)
\end{align}
From this expression, one can foresee that the Binomial distribution will be a close approximation to a Poisson distribution when the ratio $\frac{\left\langle N_{\rm sat}\right\rangle}{N}$ is small, i.e. when the number of trials is large compared to the expected number of satellites. This is actually known as the \textit{law of rare events}, and at fixed $N*p$ the Binomial distribution ${\rm Binomial}(N,p)$ converges to a Poisson distribution when the number of trials $N \rightarrow \infty$. 

\begin{figure}
    \centering
    \includegraphics[width=0.45\textwidth]{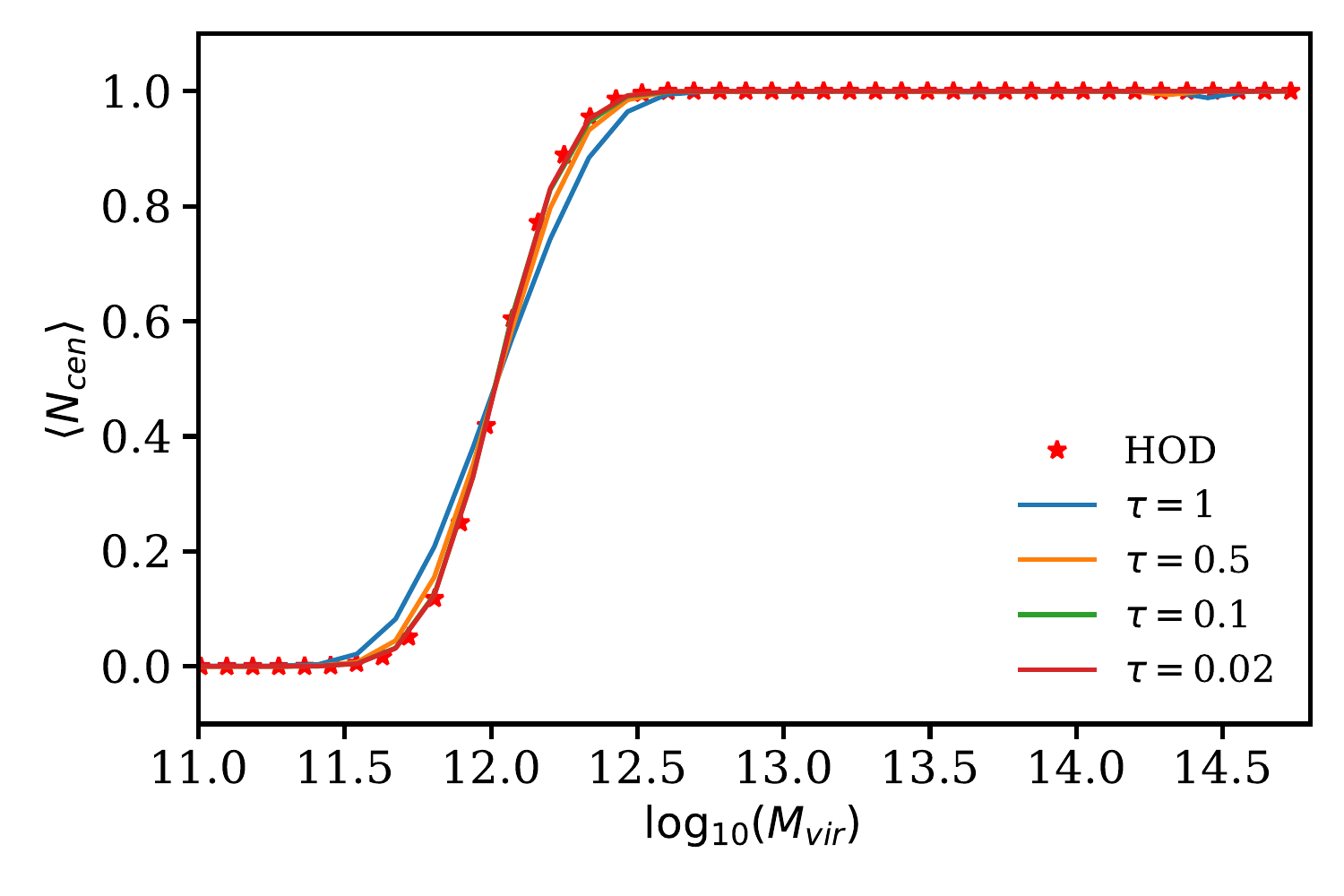}
    \includegraphics[width=0.45\textwidth]{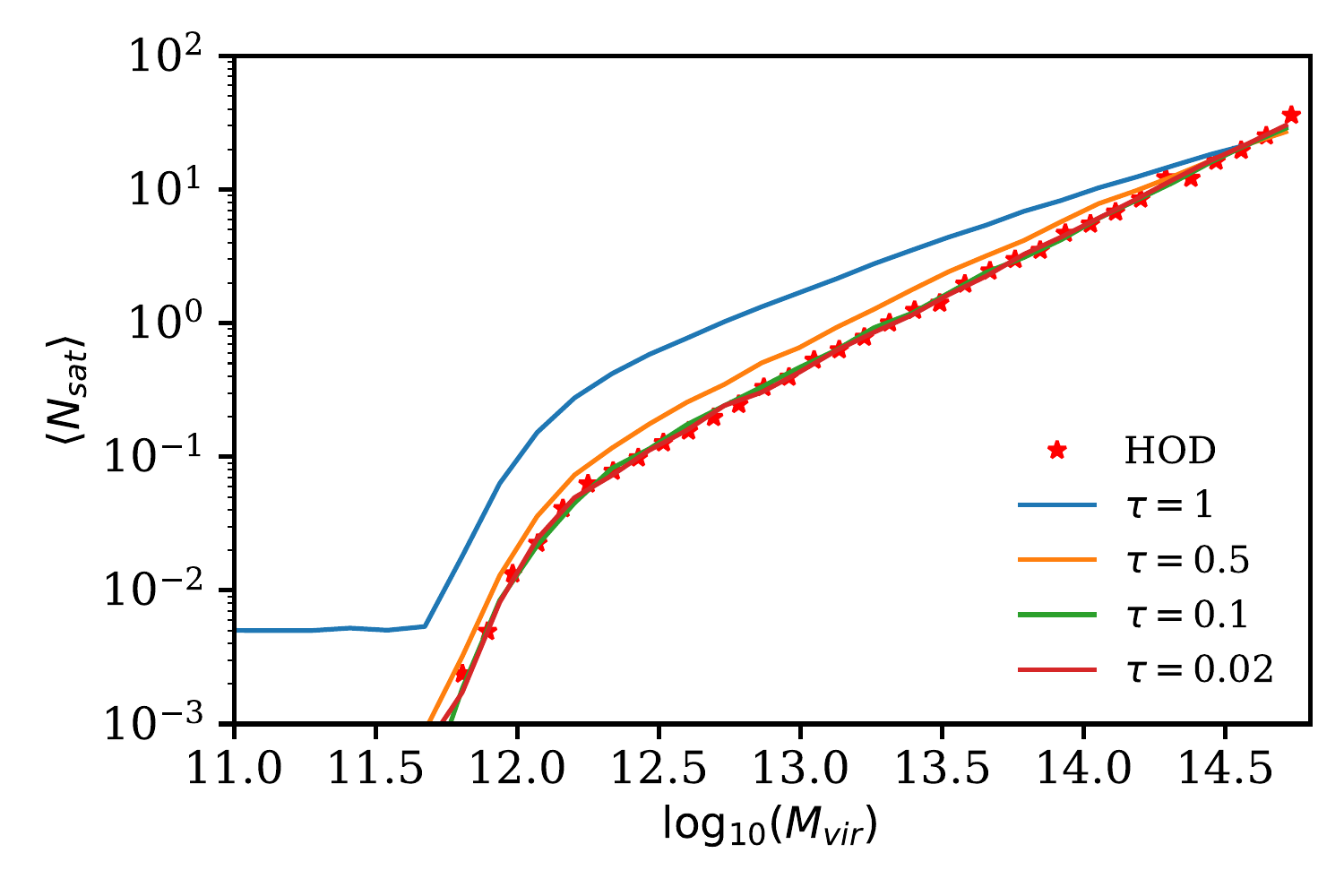}
    \caption{
    Halo occupancy distribution for central (top) and satellite galaxies (bottom) 
    as a function of halo virial mass of our differentiable HOD model (\dhod\ solid).
    Different colors indicate different temperature values used in the Gumbel-Softmax 
    approximation (see Eq. \ref{eq:softmax}).
    We include the occupancy distribution from the standard \citet{2007zheng} HOD model for 
    reference (star). 
    In this work, we use \dhod\~with $\tau=0.1$, which is in good agreement 
    with the standard HOD model throughout the full halo mass range. \nblink{Plots_for_Paper}}
    \label{fig:occupancy}
    
\end{figure}

\begin{figure}
    \centering
    \includegraphics[width=\columnwidth]{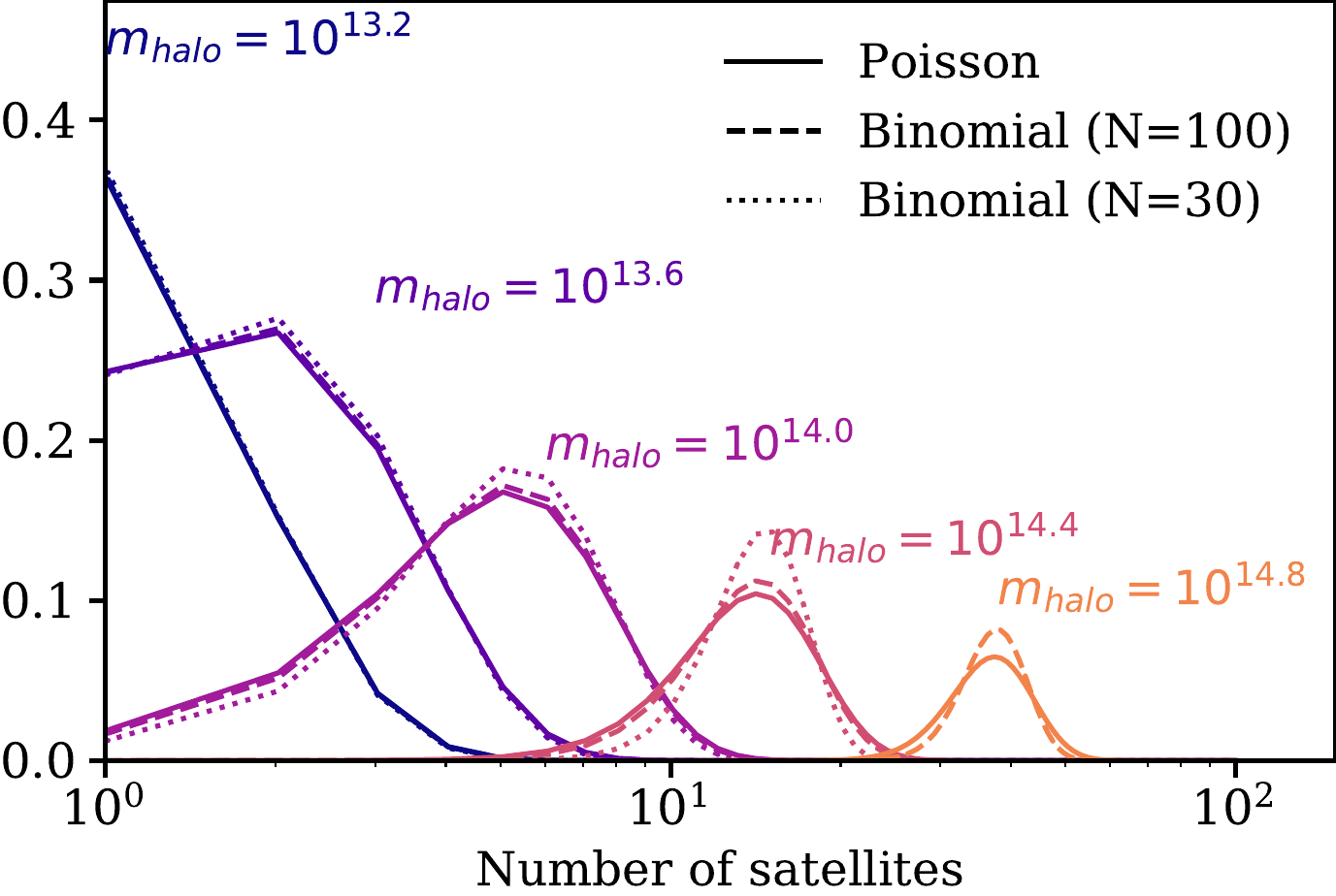}
    \caption{ Comparison of satellite occupation distributions for different halo masses, under various assumed distributions: Poisson (solid line), Binomial with 100 trials (dashed line), Binomial with 30 trials (dotted line). By construction, the Binomial approximation recovers the mean number of satellites, but for massive halos limiting the number of samples N will lead to departure in the spread of the distribution compared to a Poisson distribution. Note that for the $m_{\rm halo} = 10^{14.8}$ halo the mean number of satellites is above $30$, so the Binomial distribution is not well defined. \nblink{PoissonVSBinomial} }
    \label{fig:poisson_vs_binomial}
\end{figure}

In practice, we will need to limit the number of trials $N$ to some finite value and Fig.~\ref{fig:poisson_vs_binomial} compares the shapes of satellites distributions with a Poisson model versus a Binomial model for two different choices of $N$, and for different halo masses. A higher value of $N$ can improve accuracy but will also increases memory costs, which scales linearly with the maximum number of satellite galaxies encoded in our one-hot embedding. Meanwhile, too low of a $N$ can bias results by artificially reducing the variance of the satellite occupation distribution, or worse, truncating the satellite galaxies of the most massive halos. 

Our fiducial choice in this work is $N = 48$. In Fig.~\ref{fig:occupancy} (bottom), we can see that the satellite population reaches a maximum of $\sim 40$ galaxies for the most massive $\sim 10^{15} M_\odot$ halos. For the mass range considered here, we find $N > 40$ does not significantly improve the statistical match in summary statistics (correlation function, power spectra, etc.) of the resulting galaxy fields and larger $N$ further will increase memory requirements and computational time.
If the end statistic of interest is particularly sensitive to galaxy populations in the most massive clusters, a higher $N$ might be needed. If one is limited by memory or implementing HOD for halos with a broader mass range, then it may be more efficient to have multiple
halo mass bins with different maximum number of allowed satellites $N$.

\bigskip

This Binomial assumption for the sampling of satellites brings two concrete advantages:
\begin{enumerate}
    \item Having restated satellite sampling as draws from Bernoulli distributions, we can make the procedure differentiable by using the Relaxed Bernoulli, similarly to centrals.
    \item Using a fixed number $N$ of potential satellites gives us a practical way to handle varying number of satellites per halos. 
\end{enumerate}
Concretely, for each candidate satellite $i \in \llbracket 1,N \rrbracket$ of a halo, we sample whether the satellite will be included in the halo using:
\begin{equation}
     z_i = \frac{1}{1 + \exp( - (\log(\frac{p}{1 - p}) + \epsilon_i)/\tau) } \mbox{ with } \epsilon \sim \mathrm{Logistic}(0,1) \;.   
\end{equation}
where $p = \frac{ \left\langle N_{sat}\right\rangle }{N}$ and $\tau$ is the temperature. As a result, for each halo we obtain a vector $\mathbf{z}$ of size $N$ which encodes active satellites for the halo.

In all downstream computations, this vector $\mathbf{z}$ can be interpreted as a weight between 0 and 1 to apply to each of the $N$ satellites of each halo, for instance in the computation of two-point correlation functions. 

\subsubsection{Differentiable sampling from NFW distribution}

The last step to complete our HOD implementation is to sample the position of satellites based on an NFW profile centered at the halo position. There are two difficulties here: 1. sampling positions for a varying number of satellites, 2. making the sampled positions differentiable with respect to the NFW parameters. 

The first question of dealing with varying number of satellites is solved by our Binomial model with fixed number of trials $N$. For each halo, we will sample the same number of $N$ sets of coordinates, one for each potential satellite. Whether these coordinates will actually contribute in downstream computations will depend on the per halo satellite occupation vector $\mathbf{z}$ introduced above.

The second question, of the differentiability of stochastic coordinates, is again solved by applying the re-parameterisation trick to the NFW profile. From Eq.~(\ref{eq:NFW}) we know the CDF of the NFW profile, meaning that for each satellite $i \in \llbracket 1,N \rrbracket$ of a given halo we can sample the halo-centric satellite radial distance as:
\begin{equation}
    r_i = r_{\rm vir} \  q(\epsilon, c) \mbox{ with } \epsilon \sim \mathrm{Uniform}(0, 1)
\end{equation}
where $q$ is a differentiable function of parameters $c$. 

In our adaptation of the \cite{2007zheng} model, we assume an isotropic NFW distribution for satellite, so to retrieve halo-centric cartesian coordinates $x_i,y_i,z_i$ of a given satellite, we first sample $x_i,y_i,z_i$ on the unit sphere and then multiply these coordinates by the $r_i$ value sampled above. This is nothing more than an another reparameterisation step and the resulting cartesian coordinates remain fully differentiable with respect to the NFW parameters.

\bigskip

We note that contrary to the sampling of central of satellites which are differentiable approximation to a standard HOD (due to the discrete variables involved), this differentiable implementation of NFW sampling is exact.

\subsubsection{Impact of temperature parameter $\tau$}
In the \dhod~model, we introduce temperature, $\tau$, as a 
free parameter (Eq.~\ref{eq:diff_cen}).
Depending on the context/implementation, it may be beneficial to anneal 
(\emph{i.e.} reduce) $\tau$ over the course of the optimization such that $\tau \rightarrow 0$. For instance, in the original papers \citep{2016arXiv161101144J,2016arXiv161100712M}, the Gumbel-Softmax trick was used in the context of training a neural network where only the optimal network weights were of interest and hence $\tau$ was reduced to nearly zero over the course of the training.
However in our case, we are interested in stochastic sampling, and not 
optimization, so we will pick a single temperature that well approximates 
the target distribution (\emph{i.e.} unbiased) while maintaining reasonable 
derivative properties (i.e. less noisy). 

\begin{figure*}
    \centering
    \includegraphics[width=0.70\textwidth]{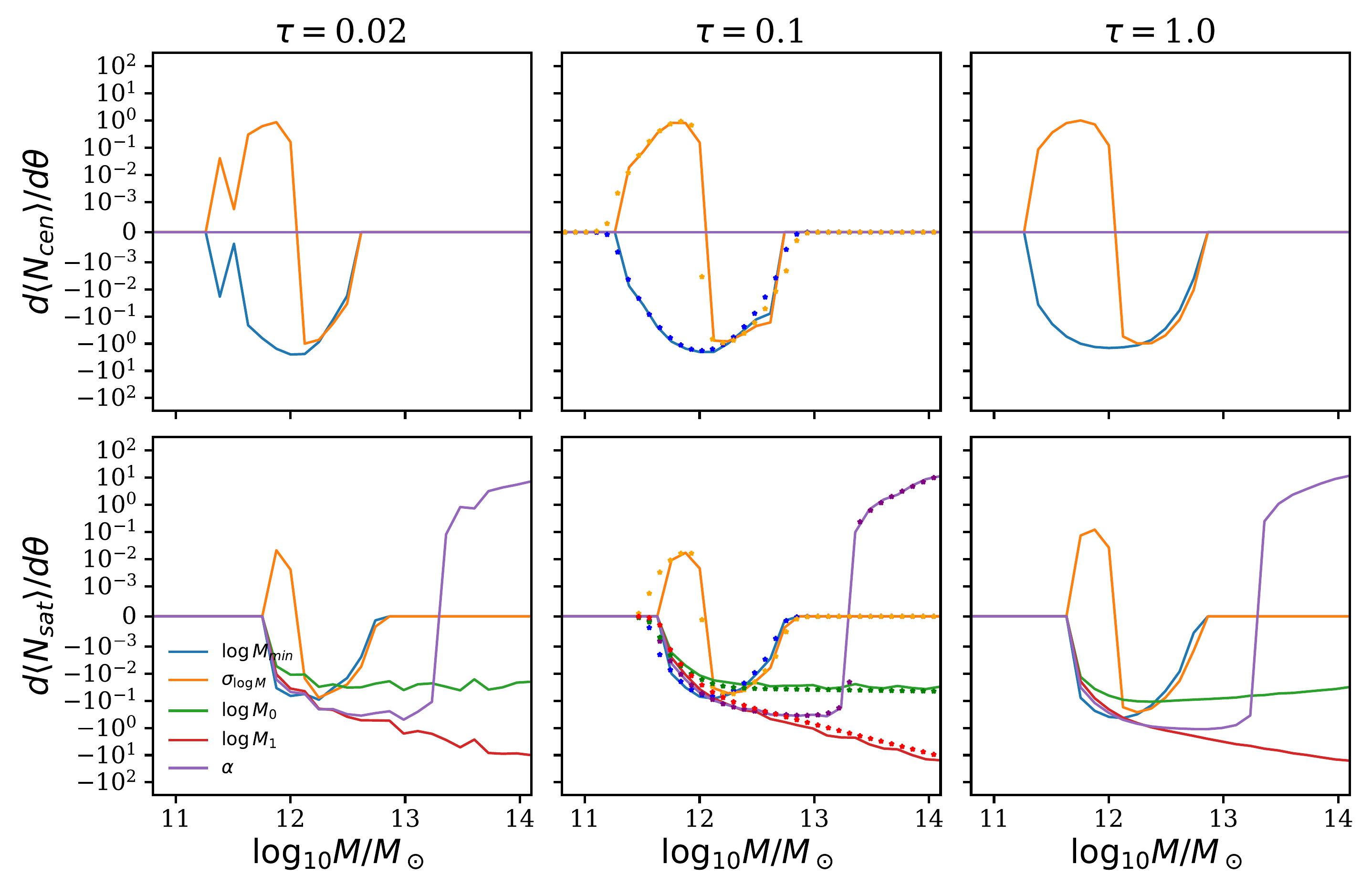}
    \includegraphics[width=0.71\textwidth]{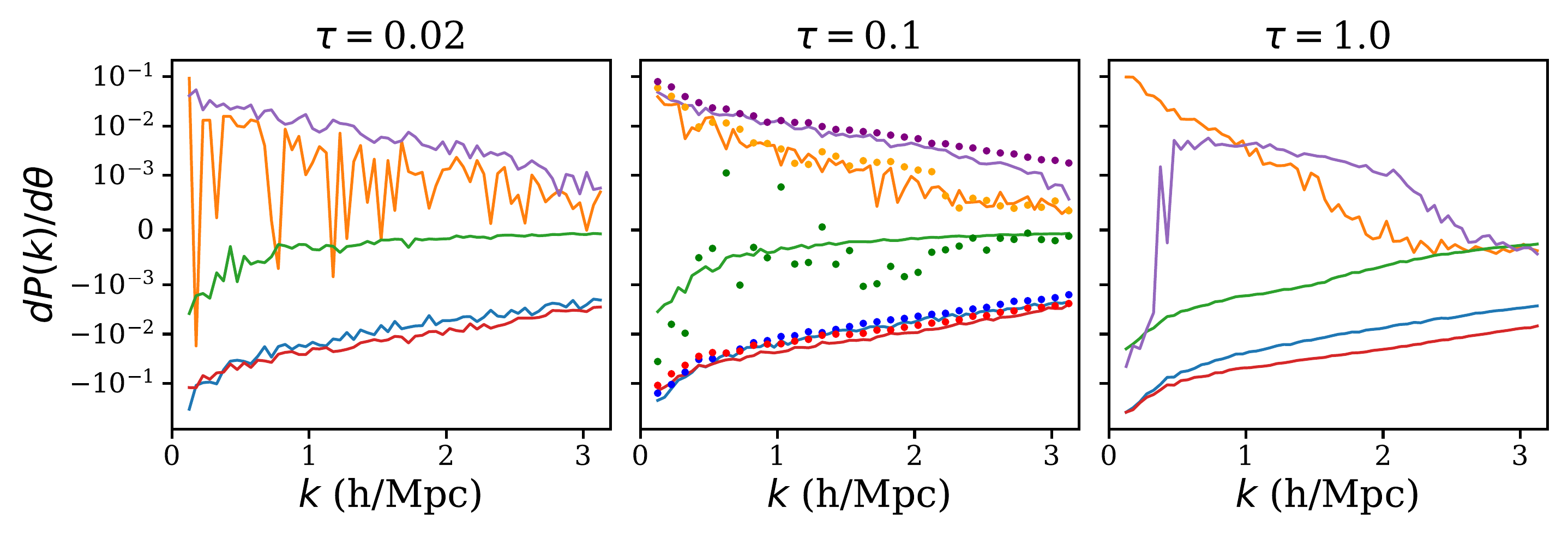}
    \caption{
    \emph{Top}: Derivatives of the central and satellite occupancy functions 
    with respect to the HOD parameters for various temperatures: $\tau=0.02$ 
    (left) 0.1 (center), and 1 (right). 
    The derivatives are calculated at the fiducial HOD parameter values. 
    We include derivatives analytically derived from Eqs.~\ref{eq:ncen} 
    and~\ref{eq:nsat} for comparison (star). 
    \emph{Bottom}: Derivatives of the galaxy power spectrum with respect to the 
    HOD parameters, evaluated at the fiducial HOD parameter values. 
    We include derivatives calculated using the standard HOD with finite 
    difference for comparison (star). 
    Our \dhod~model with $\tau=0.1$ provides sufficiently smooth derivatives that
    are in good agreement with analytical derivatives for the occupancy function 
    and with standard HOD derivatives for the power spectrum. \nblink{Plots_for_Paper}
    %Small dots represent derives calculated analytically from Eq. \ref{eq:ncen} and Eq. \ref{eq:nsat} (top) and finite difference (bottom).
    \label{fig:occupancy_derivatives}}
\end{figure*}

%We show in Figure \ref{fig:occupancy} how the occupancy distributions change as a function of temperature compared to the that sampled from halotools.
In Fig.~\ref{fig:occupancy}, we present the central (top) and 
satellite (bottom) occupancy distributions of our \dhod~model
for different temperatures: 
$\tau = 0.02$ (red) 0.1 (green), 0.5 (orange), and 1 (blue). 
We include the occupancy distributions for the standard HOD model for
comparison (star).
For this work, we take an experimental approach for determining
    $\tau$, as advocated in \citet{2016arXiv161100712M}:
    the temperature should set as high as possible while maintaining the 
    desired accuracy of the target distribution. 
    We find that a fixed $\tau=0.1$ provides high accuracy 
    while maintaining stable gradients for both sampling the number of 
    galaxies in each halo as well as the positions of the satellites from 
    the NFW profile (see later Fig.~\ref{fig:occupancy_derivatives}).

\section{Experimentation}
\label{sec:exp}

%\CH{I think we should switch many of the references to halotool tools to just ``the standard''. 
%Halotools isn't some seminal package in the field so I think this will be confusing to people.}\BH{Happy to change the nomenclature! It might be nice to have a paragraph summarizing the different implementations and why we choose to use halotools for our comparison.}

To test our implementation, we construct a fiducial mock galaxy catalog from the Planck Bolshoi simulation halo catalog
\citep{2011Bolshoi} at $z=0$. We treat this catalog as our mock observation. This simulation has a  side length of 250 $h^{-1}$ Mpc, 
and contains 1,367,493 unique halos ranging in mass from $1.1 \times 10^{15} M_\odot$ down to $2.7 \times 10^{8} M_\odot$. We use halotools \citep{2016MNRAS.460.2552H} to generate our fiducial catalog with  HOD parameter values: 
\begin{gather*}
    \log{M_\textrm{min}}=12.02,  \sigma_{\log{M}}=0.26, \log M_0 = 11.38, \\  \log{M_1}=13.31, \alpha = 1.06.
\end{gather*}
These parameters correspond to the best fit values from \citet{2007zheng} using an SDSS-based galaxy catalog \citep{2005ApJ...630....1Z} with $r$-band absolute magnitude threshold of $M_r < -20$.

\subsection{Summary Statistics and Derivatives with \dhod}

Next, we compare the mock observations to parallel galaxy catalogs constructed using \dhod~(Section~\ref{sec:methods}). 
The galaxy catalogs are then painted onto a grid using a differentiable Cloud-In-Cell (CIC) painting method \citep{2020arXiv201011847M} and its real-space power-spectra calculated via a differentiable Tensorflow power spectrum implementation \citep{2021TARDISII}.
For \dhod, all steps are differentiable so the overall mapping from 
original halo catalog to end power spectrum is also differentiable via the
chain rule. 
In Fig.~\ref{fig:occupancy_derivatives}, we present the \dhod~derivatives 
of the central and satellite occupancy distribution functions (top)
and resulting power spectra (bottom) with respect to the HOD parameters
for different temperatures: $\tau = 0.02$ (left), $0.1$ (middle), and $1$ (right). 
The derivatives are evaluated at the fiducial HOD parameter values. 
In the center panels, for comparison, we include derivatives of the central 
and satellite  occupancy distribution functions derived analytically and 
derivatives of  the power spectrum derived using the standard HOD with finite 
differences (star). 

With $\tau=0.02$, the power spectrum derivatives have significant numerical noise. 
The $\tau = 1.0$ derivatives are smoother but we find inaccurate occupancy
distributions (Fig.~\ref{fig:occupancy}) and a significantly biased power 
spectrum.
Meanwhile, with $\tau = 0.1$ there is still some noticeable numerical noise
in the derivatives, however, this is sufficiently smooth for our application 
and for our optimization to be well behaved. 
We also find that the $\tau=0.1$ \dhod~derivatives are in good agreement with
analytical derivatives for the occupancy function and with standard HOD derivatives 
estimated using finite differences for the power spectrum.

%We find the derivatives from this process match well those found via analytical calculation of the HOD model or those from finite differences of the resulting power spectrum. 
%\FL{Did I miss it or are we not mentioning anywhere that we are actually with non perfect derivatives because of the metropolis hastings step in the HMC?}
%\CM{Not sure what you mean}

\begin{figure}
    \centering
    \includegraphics[width=0.50\textwidth]{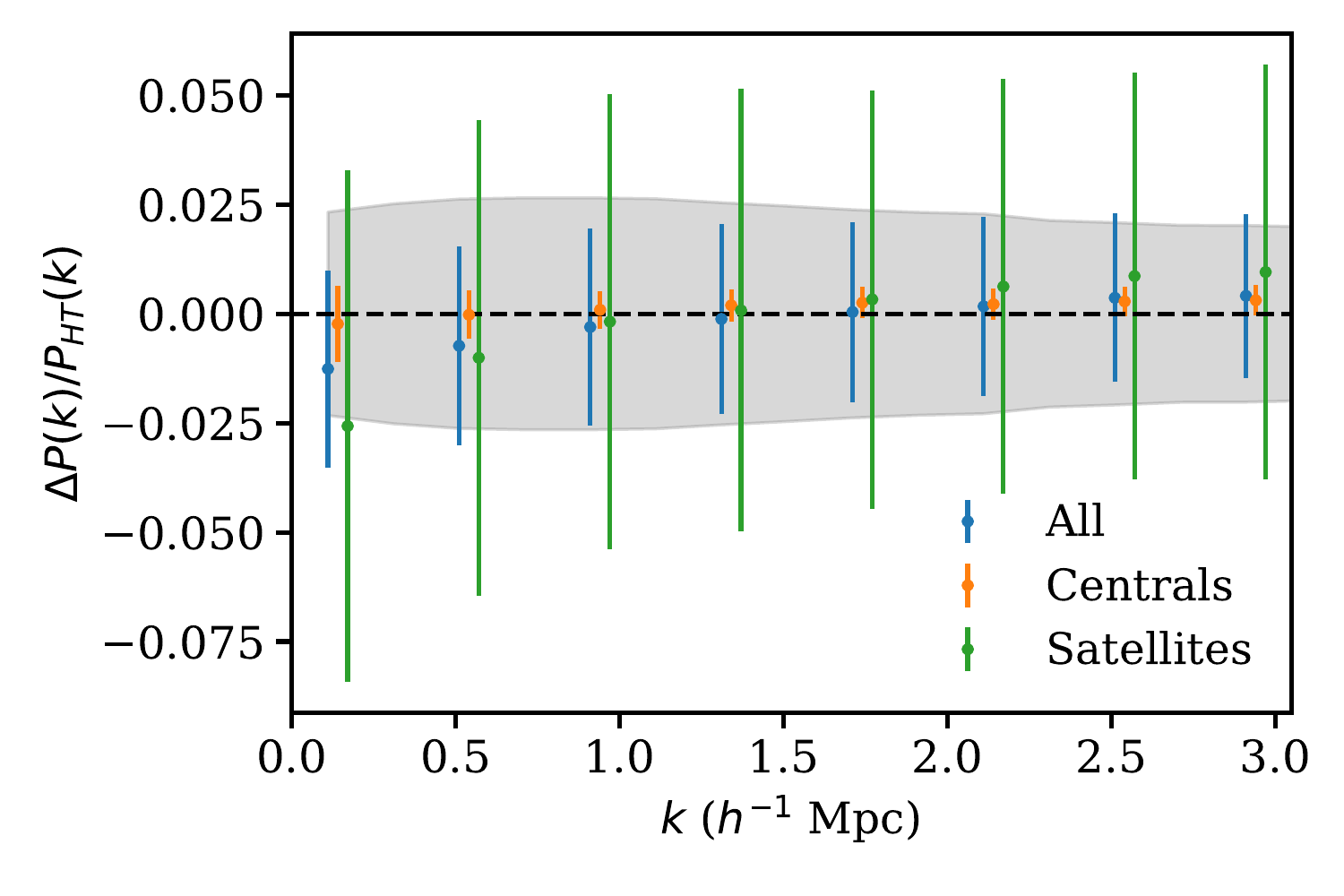}
    \caption{
    Comparison of power spectrum derived using a standard HOD model ($P_{\rm HT}$)
    versus our \dhod~model. 
    We plot the ratios for central and satellite galaxies, as well as for all galaxies. 
    Error bars represent the standard deviation of the \dhod~model estimated from 
    500 realizations of the galaxy sampling. Grey band represents the standard deviation of $P_{\rm HT}$ for all galaxies.
    We find good agreement between the differentiable power spectrum and $P_{\rm HT}$ 
    across all scales ($k < 3 h/{\rm Mpc}$), as well as similar distributional properties. \nblink{power_spectra_comparison}
    \label{fig:ps}}
\end{figure}

We compare the differentiable power spectrum from \dhod~to the power spectrum 
from the standard HOD model in Fig.~\ref{fig:ps}.
We use the fiducial HOD parameter values and estimate the error bars from 500 
realizations of the \dhod~model. 
We find good agreement (better than $1\%$) across all scales, with particularly
good agreement for central galaxies, which are not sampled from the NFW profile. 
Low $k$ modes are most sensitive to the most massive halos whose satellite 
galaxy populations are truncated by our one-hot distribution 
($N=48$; Section~\ref{sec:dhod}); those who are interested in this regime can further increase $N$.
However, even at $k < 0.15 h/{\rm Mpc}$, we find good agreement between the power spectra.
    
\subsection{Monte Carlo Analysis with \dhod}

Lastly, we demonstrate that we can derive unbiased inference using \dhod, by
    comparing the posteriors on HOD parameters derived using \dhod~to posteriors
    derived using standard methods with the same mock observations and likelihood. 
    We use the power spectrum measured from the fiducial galaxy catalog as our mock
    observations, and construct a covariance matrix using 100 galaxy catalog
    realizations at the fiducial HOD catalogs. To this covariance, we added a small constant diagonal term ($8.0 \times 10^{-5}$) to improve numerical stability. We limit our comparison to $k<1.0$.
    In both \dhod~and standard cases, we use the same halo catalog used to construct
    the mock observations, so that the only source of error is variation caused by
    the HOD model. For this analysis we impose wide Gaussian priors, $\mathcal{N}(\mu,\sigma^2)$, where $\mu$ is the mean, and $\sigma^2$ is the variance, on our parameter values as follows: 
\begin{gather*}
    \log{M_\textrm{min}} \sim \mathcal{N}(12.0,0.5),  \sigma_{\log{M}}\sim \mathcal{N}(0.25,0.2), \\\log M_0 \sim \mathcal{N}(11.25,0.5), \log{M_1} \sim \mathcal{N}(13.20,0.5),  \\ \alpha \sim \mathcal{N}(1.0,0.2).
\end{gather*}
%We now want to find the posterior on HOD parameters based on the observed power-spectra of the mock galaxy catalogs. We optimize for the HOD parameters within our differentiable forward model.
%For demonstration, we use the same halo catalog as used to construct the mock catalog so the only source of error is variation caused by the HOD model. We construct a covariance matrix for our power spectra around a fiducial HOD parameter set for use in our likelihood calculation.
To derive the posteriors using \dhod, we sample over the HOD parameter using 
Hamiltonian Monte Carlo (HMC) \citep{duane1987hybrid,neal2011mcmc}. We use the NoUTurn HMC implementation \citep{hoffman2014no} in Tensorflow Probability.
We use three chains initialized around our fiducial HOD parameters with over 1,000 steps (300 steps of burn-in).
    
\begin{figure}
    \centering
    \includegraphics[width=0.50\textwidth]{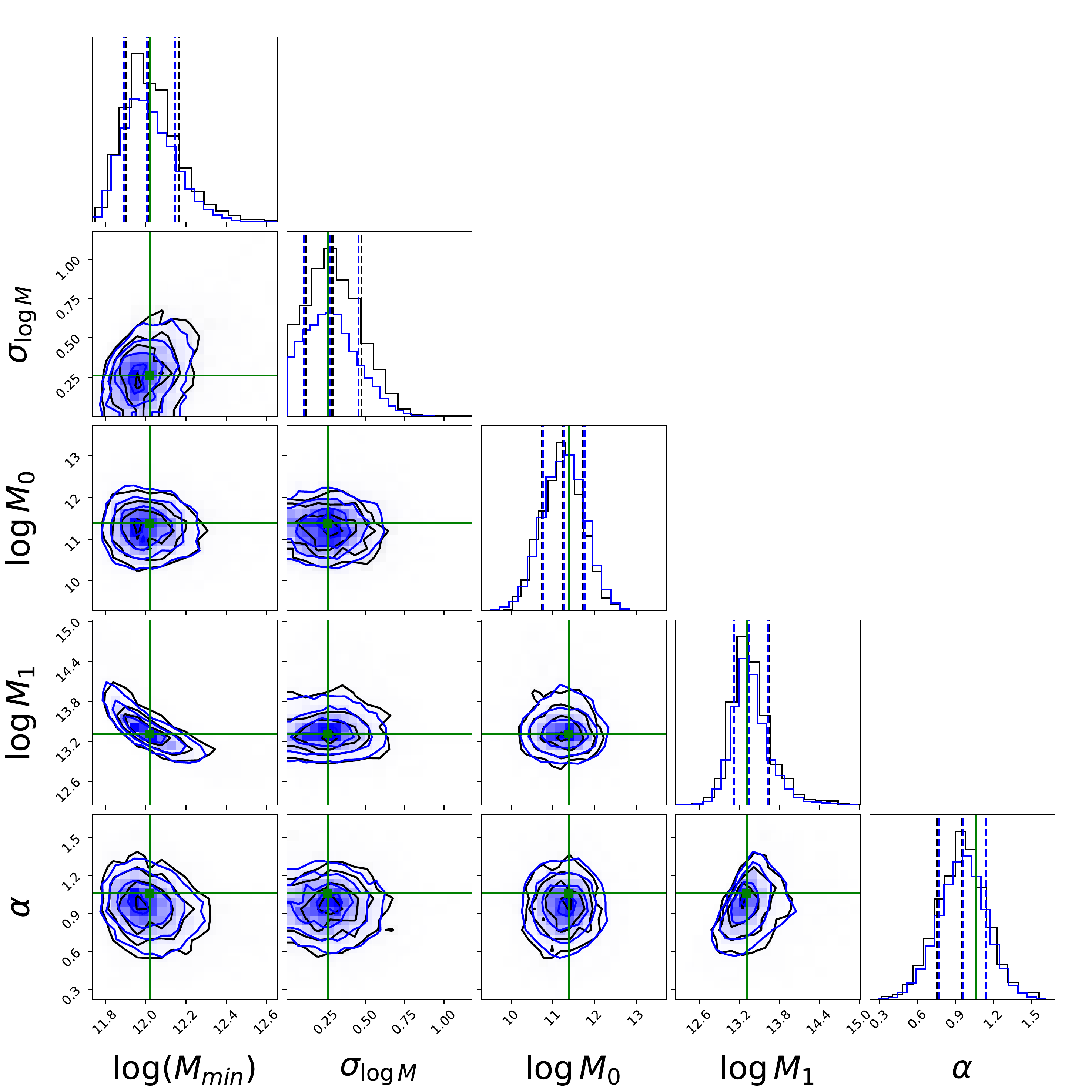}
    \caption{Posterior distributions of the \citet{2007zheng} HOD parameters derived 
    using \dhod~with Hamiltonian Monte Carlo (black) and the standard HOD with Markov
    Chain Monte Carlo (blue).
    We use a single mock galaxy catalog realization at the fiducial HOD parameter 
    values (green) as our observations. 
    %We derive $4\times$ faster than standard HOD analyses with our \dhod~and HMC.
    \nblink{chain_scripts/chain_analysis_5d}
    }
    %Output of the Hamiltonian Monte Carlo chain from the DiffHOD model (Red) and Markov Chain Monte Carlo chain from the HOD analysis (Black) for the mock dataset. }
    \label{fig:hmc}
\end{figure}

For the standard approach, we run the same analysis using the standard HOD. 
However, since this implementation does not allow easy differentiation,
we cannot use HMC instead use a Markov-Chain Monte Carlo analysis. 
We use the emcee \citep{2013PASP..125..306F} implementation with 10 walkers 
and 6,000 steps.
We present the posteriors on the HOD parameters for the \dhod~(black) and 
    standard (blue) analyses in Fig.~\ref{fig:hmc}
and list the median posterior values in Table~\ref{table:values}. 
We mark the fiducial (``true'') HOD values in green. 
As we are using a single galaxy realization for our mock observations, 
we expect some variation between the best fit parameters and the true values.
 The posteriors derived using \dhod~and HMC is in excellent agreement with 
    the  posteriors derived using the standard HOD and MCMC.

Our HMC analysis takes approximately 10 hours on a single Tesla V100-PCIE-32Gb GPU. 
Meanwhile, the standard approach takes substantially more
time to get comparable results: $\sim$200 hours on 1 CPU --- ${\sim}20\times$ 
slower than our \dhod~analysis. 
Some of this improvement is due to the fact that our \dhod~implementation is
faster per iteration than the standard HOD implementation (${\sim}1$ and 
${\sim}4$ seconds per iteration, respectively).
Most of the improvement, however, comes from the fact that \dhod~allows us
    to exploit a more efficient gradient-based method to derive the posterior. 

%We note that the \dhod~implementation is faster per iteration (${\sim}1$ second) than the standard HOD implementation (${\sim}4$ seconds), which accounts for some of the improvement. 
%accounting for some of the relative improvement.  \FL{Ok, but does that mean we are comparing against running diffhod with emcee?}

We compare the \dhod~and standard approaches in more detail by comparing the
effective sample size of each
%A more nuanced analysis can be performed by comparing the effective sample size of each 
chain per function evaluation \citep{gelman2013bayesian}. 
The effective sample size incorporates information about the auto-correlations 
within a chain; \emph{i.e.} it accounts for the dependent relationships 
between the samples. 
We calculate it from the output of the Markov chain:
\begin{equation}
    N_\textrm{eff} = \frac{N}{1+\sum\limits_{t=1}^{\infty} \rho_t}
\end{equation}
where $N$ is the number of samples in the chain and $\rho_t$ is the autocorrelation
of length $t$.
Averaging over all HOD parameters, we find an mean effective sample size of $524.2$ for 
our \dhod~HMC evaluation and $403.9$ for the standard MCMC evaluation. 
This corresponds to an effective sample of 0.05 per step for the HMC and 0.006 for the MCMC evaluation.

\section{Conclusions}
\label{sec:conclusion}
In this work we have constructed a differentiable stochastic HOD model going from a halo catalog to an observed galaxy power spectrum.
This allows us to use derivative-based optimization methods to quickly optimize for the underlying model parameters. 
This is the first time differentiable stochastic models have been used in the astrophysics literature. 
We find the \dhod~model provides a 
$4\times$ increase of speed versus the same analysis performed via MCMC with the standard HOD implementation. 
\dhod~is an alternative to a number of recent works focusing on emulating galaxy clustering statistics \citep{2015ApJ...810...35K,2019MNRAS.484..989W,2020PhRvD.102f3504K,2020MNRAS.492.2872W, Hearin2022}. 
Unlike emulator based methods, \dhod~works at the level of the halo catalog and allows fast, differentiable, generation of any summary statistic with respect to the HOD parameters.

In this work, we have focused on the \citet{2007zheng} HOD model, but our methods can be easily extended to a broad class of models. While standard HODs are based only on halo mass, in general various properties of the halos environment and formation history could effect the galaxy properties \citep{2006ApJ...639L...5Z,2007MNRAS.374.1303C}. Galaxy assembly bias has been argued \citep{2015MNRAS.446.1939F,2020MNRAS.493.5506H} to cause significant deviations between predictions of standard HOD models and those from hydrodynamical simulations. Decorated HOD models have been introduced to account for assembly bias \citep{2016MNRAS.460.2552H} and has been extended to include other possible effects \citep{2018MNRAS.478.2019Y}. These models still rely on stochastic discrete sampling for assigning centrals and satellites so they can be modelled in a differentiable way using the techniques described in this work. As the dimensionality of our problem increases, either with extended HOD models or joint analysis with cosmological parameters, we expect the relative performance of derivative based methods, like HMC, over pure sampling based methods to further improve \citep{neal2011mcmc}.

\begin{table}
\centering
\def\arraystretch{1.5}
     \begin{tabular}{ c | c | c }
 \hline\hline  
     & \dhod & Standard HOD \\ \hline
    $\log(M_\textrm{min})$ & $12.03^{+0.15}_{-0.03}$ & $12.01^{+0.14}_{-0.02}$ 
\\
    $\sigma_{\log{M}}$ & $0.28^{+0.19}_{-0.16}$ & $0.27^{+0.18}_{-0.16}$
\\
    $\log M_0$ & $11.25^{+0.38}_{-0.43}$ & $11.27^{+0.49}_{-0.50}$ 
\\
    $\log{M_1}$ & $13.32^{+0.23}_{-0.23}$ & $13.34^{+0.29}_{-0.22}$
\\
 $\alpha$ & $0.96^{+0.19}_{-0.19}$ & $0.96^{+0.18}_{-0.19}$
\\
  \hline
 \end{tabular}
 \caption{Posterior values from the HOD analyses using the \dhod~and standard 
 HOD model. Uncertainties are estimate from the $16\%$ and $84\%$ quantiles.} 
 \label{table:values}
\end{table}
Differentiable HOD models have even more apparent applications in the case of dynamical forward model large scale reconstructions \citep{2017JCAP...12..009S} when paired with efficient differentiable halo finding methods \citep{2018JCAP...10..028M,2020arXiv201011847M,2019PhRvD.100d3515K}. While it is possible to perform these reconstructions by interpreting the galaxy field as a biased version of the dark matter field (i.e. in \cite{2021TARDISII}), inaccuracies in this prescription will result in biases that would be difficult to account for in cosmological constraints. Through joint inference of the HOD parameters with the initial density field, these astrophysical uncertainties can be rigorously marginalized out. Differentiable models are critical for this application as the optimization is highly multidimensional (approximately number of particles in the simulation) and would be computationally infeasible without gradient-based methods.

While in this work we have highlighted using our \dhod~model inside an HMC framework, one can exploit
its automatic differentiation for a variety of first order optimization and parameter inference methods. For example standard Variational Inference relies on having well defined derivatives for the optimization of latent space parameters describing the likelihood surface \citep{peterson1987mean,2003PhDT.......250B,2016arXiv160100670B}. Variational inference could further accelerate parameter inference when compared to Hamiltonian Monte Carlo or nested sampling methods \citep{2018arXiv180306473G}.

\section*{Acknowledgements}
We thank Andrew Hearin and Matt Becker for very useful discussions and suggestions.
BH and CH are supported by the AI Accelerator program of the Schmidt Futures Foundation. 
SF is supported by the Physics Division of Lawrence Berkeley National Laboratory.
This research used resources of the National Energy Research Scientific Computing Center, a DOE Office of Science User Facility supported by the Office of Science of the U.S. Department of Energy under Contract No. DEC02-05CH11231.

%%%%%%%%%%%%%%%%%%%%%%%%%%%%%%%%%%%%%%%%%%%%%%%%%%



\bibliographystyle{mnras}
\bibliography{ref}

%%%%%%%%%%%%%%%%%%%%%%%%%%%%%%%%%%%%%%%%%%%%%%%%%%

%%%%%%%%%%%%%%%%% APPENDICES %%%%%%%%%%%%%%%%%%%%%
\appendix
\section{Distributional Properties of \dhod\ Model}
\label{ap:1}
\begin{figure*}
    \centering
    \includegraphics[width=0.90\textwidth]{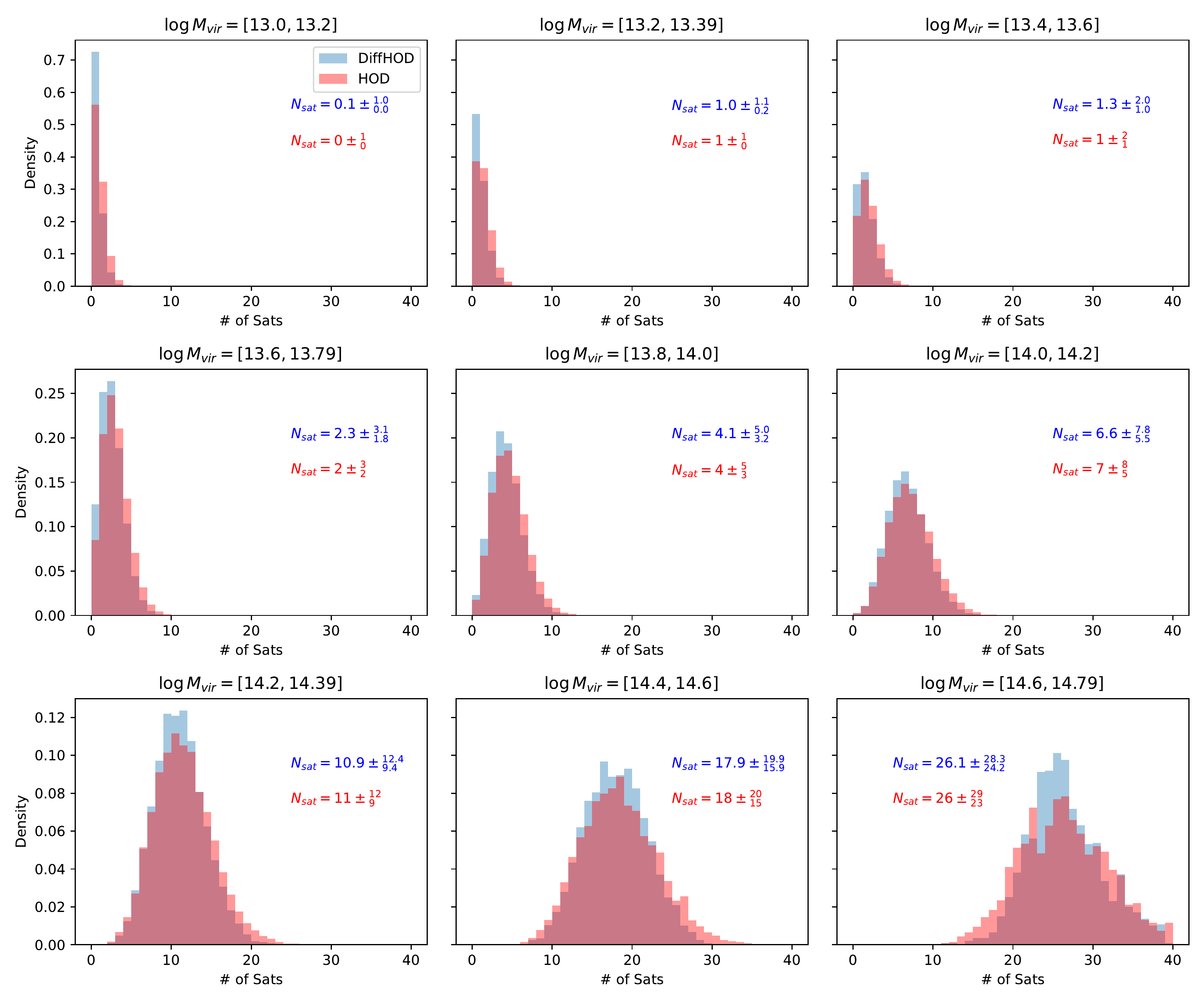}
    \caption{Histogram showing the halo occupancy distribution from 100 independent samples from the complete halo catalog. We compare halos populated by our \dhod model as well as a standard HOD implementation. We show the median satellite occupation number, and quote error bars representing the 32\% and 68\% quartiles. We find excellent quantitative and qualitative agreement between the two distributions. 
    \nblink{Plots_for_Paper}}
    \label{fig:distribution}
\end{figure*}

In the main body of the work we sampled from a Bernoulli Distribution rather than a pure Poisson Distribution due to the existing analytical tools to relax the Bernoulli Distribution via the Gumbel-Softmax trick. This was demonstrated to be a valid approximation at the level of various summary statistics, such as halo occupancy distribution functions and resulting galaxy power spectrum. In this section we show the halo occupancy distribution as a function of halo mass.

We sample our satellite \dhod, at $\tau=0.1$, and the standard satellite HOD using the parameters in the main text 100 times in order to attain reasonable number statistics at the high mass bins. We show our results in \ref{fig:distribution}, finding quantitative good agreement between the models as calculated by their modal and variance properties. We calculate for each mass bin the 32\%, 50\% and 68\% percentiles. Since \dhod\ uses a relaxed distribution instead of sampling, it is possible to get non-integer number of satellites while for the standard HOD model all sampling is discretized. Qualitatively we see a slight broadening of the distributions at the extreme high mass end, however this does not noticeably impact any of our resulting analysis due to the very small population of these extreme high mass halos. Additional optimizations in terms of maximum satellite population and choice of temperature could be performed if this populations is of high interest.

%%%%%%%%%%%%%%%%%%%%%%%%%%%%%%%%%%%%%%%%%%%%%%%%%%

% Don't change these lines
\bsp	% typesetting comment
\label{lastpage}
\end{document}